\documentclass[aps,prl,twocolumn,showpacs]{revtex4}
\usepackage{amssymb}
\usepackage{graphicx,amsmath}

\newcommand{\bea}{\begin{eqnarray}}
\newcommand{\eea}{\end{eqnarray}}
\newcommand{\beq}{\begin{equation}}
\newcommand{\eeq}{\end{equation}}
\newcommand{\benu}{\begin{enumerate}}
\newcommand{\enu}{\end{enumerate}}

\newcommand{\al}{\alpha}

\newcommand{\om}{\omega}
\newcommand{\Om}{\Omega}
\newcommand{\ep}{\epsilon}

\newcommand{\si}{\sigma}

\newcommand{\ptl}{\partial}

\newcommand{\bk}{{\bf k}}
\newcommand{\bq}{{\bf q}}
\newcommand{\bp}{{\bf p}}

\newcommand{\bv}{{\bf v}}
\newcommand{\bE}{{\bf E}}
\newcommand{\bL}{{\bf \Lambda}}
\newcommand{\bJ}{{\bf J}}

\begin{document}

\title{
Equivalence of single-particle and transport lifetimes from
hybridization fluctuations}
\date{\today}
\author{I. Paul$^1$, C. P\'epin$^2$ and M. R. Norman$^3$}
\affiliation{
$^1$Laboratoire Mat\'{e}riaux et Ph\'{e}nom\`{e}nes Quantiques, Universit\'{e} Paris Diderot-Paris 7 \& CNRS,
UMR 7162, 75205 Paris, France \\
$^2$IPhT, CEA-Saclay, L'Orme des Merisiers, 91191 Gif-sur-Yvette, France\\
$^3$Materials Science Division, Argonne National Laboratory,
Argonne, IL 60439, USA}

\begin{abstract}

Single band theories of quantum criticality successfully describe a
single-particle lifetime with non-Fermi liquid temperature
dependence. But, they fail to obtain a charge transport rate with
the same dependence unless the interaction is assumed to be momentum
independent. Here we demonstrate that a quantum critical material,
with a long range mode that transmutes electrons between light and
heavy bands, exhibits a quasi-linear temperature dependence
for {\it both} the single-particle and the charge transport
lifetimes, despite the strong momentum dependence of the
interaction.

\end{abstract}

\maketitle
\emph{Introduction.}--- One of the most fascinating and least
understood properties of the non-Fermi liquid state that is obtained
by tuning a heavy fermion metal to a quantum critical point is the
temperature dependence of its resistivity $\rho (T) \sim T^{n}$ with
$n \approx 1$~\cite{Lohneysen:2007ve,Stewart:2006bz}. This anomalous
$T$-dependence is a rather general phenomenon whose relevance
extends beyond the field of heavy fermions. The most well-known
example is perhaps the linear-$T$ resistivity observed in the
cuprates in their normal phase~\cite{leenagaosawen,normanpepin}.
Strikingly, the same behavior is found close to a metamagnetic phase
transition in Sr$_3$Ru$_2$O$_7$~\cite{AWRosta:2011vy}. While in
heavy fermion metals the behavior is sometimes sub-linear in
$T$~\cite{Paglione:2006ut} and sometimes
quasi-linear~\cite{Friedemann:2009vn}, there is now a large body of
experimental data that establishes the universality of the
observation of anomalous exponents~\cite{Coleman:2001tp}.

Despite years of research on this important issue,
at present there are very few microscopic mechanisms known which can
explain the $\rho \sim T$  behaviour seen in clean systems. While the marginal Fermi liquid theory of
Varma \emph{et al.}  developed in the context of the cuprates is phenomenologically
constructed by assuming momentum independence of the underlying scattering~\cite{varmamfl},
the microscopic proposal of Rosch~\cite{Rosch} in the context of antiferromagnetic quantum critical
points requires special dimensionality of the interaction as a phenomenological input.

In this paper we show that a quasi-linear
$T$-dependence of the resistivity is established in multi-band systems
in the case where a hybridization interaction between the bands, with fluctuations
having overdamped dynamics,
becomes long-ranged. Such a situation arises when a system with light and heavy bands,
with Fermi wavevectors similar in magnitude, gets close to a so-called
Kondo breakdown~\cite{Paul:2007th,Paul:2008bg}, or
equivalently an orbitally-selective Mott quantum critical point~\cite{Pepin:2007do}.

Any microscopic mechanism that attempts to establish a quasi-linear $T$-dependence of
$\rho$ usually needs to satisfy two requirements. First, the putative theory should have
the means to establish non-Fermi liquid characteristics in the single particle properties.
Second, the mechanism should allow the identification of the particle lifetime with the
transport time. Quantum critical theories with single bands satisfy the first but not the
second criterion in the clean limit. This is because, in theories where the critical mode is
at wavevector $q = 0$, the singular scattering
involves small momentum transfer which is ineffective for the
relaxation of the charge current~\cite{mathon}. On the other hand, if the critical mode has a finite
wavevector, the non-Fermi liquid feature is established only in the `hot' regions which are
short-circuited for charge transport~\cite{hlubinarice}.
Our main aim in this paper is to demonstrate that
in a multi-band system, even the second criterion can be fulfilled in the case where the
critical mode is at $q = 0$, provided it involves transmutation of a light electron into a heavy
one and vice versa (i.e., a hybridization fluctuation).

\begin{figure}[!!t]
\begin{center}
\includegraphics[width=5cm]{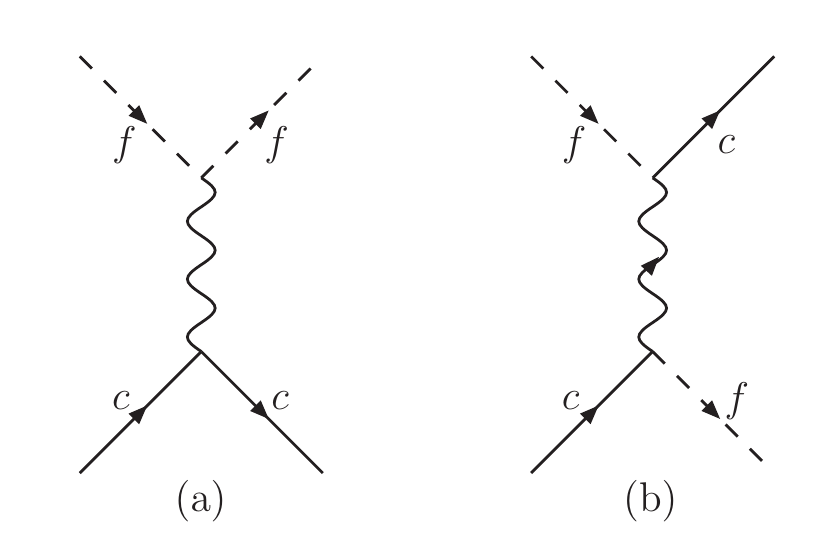}
\caption{
Interband interactions (wavy lines) between light $c$ (solid lines) and heavy $f$ (dash lines)
electrons (a) without, and (b) with band transmutation. Only in case (b), which is relevant for
the Kondo-Heisenberg model, are the particle and transport lifetimes equivalent,
leading to quasi-linear $T \log T$ resistivity.
}
\label{fig1}
\end{center}
\end{figure}

The physical reasoning underlying the result is as follows. In a minimal two-band model, with a
conduction $c$-band (light) and a correlated $f$-band (heavy),
Gallilean invariance is broken due to the inequality of the
fermion masses~\cite{appeloverhauser}. The
charge current operator is not merely proportional to the total momentum of the two bands
(in which case the
conductivity would be infinite in the absence of a lattice), but it also depends on the relative
momentum. Thus, even in the absence of a lattice (and Umklapp scattering), one can obtain finite
resistivity in a model where the part of the current proportional to the total momentum
is relaxed by impurity scattering
(giving rise to a $T$-independent contribution), and the part proportional to the relative momentum
is relaxed by
interband electron-electron scattering (giving a $T$-dependent contribution). However, having two
bands is not sufficient for establishing equivalence of particle and transport times. For example,
if the interband interaction does not involve particle transmutation as in Fig.~\ref{fig1}(a), the relative
momentum is the same before and after the scattering in the asymptotic limit of zero momentum transfer.
In this case, in the relevant $T$-regime, the transport time has additional $T$-dependence from the
$(1- \cos \theta)$ factor because the scattering angle $\theta$ of the relative momentum is
$\theta \approx 0$~\cite{maslov}.
In contrast,
if the interband interaction involves hybridization fluctuation as in Fig.~\ref{fig1}(b), the relative momentum
undergoes back-scattering in the limit of zero momentum transfer such that $\theta \approx \pi$, and there is
no additional $T$-dependence.
Thus, when such an interaction has overdamped dynamics (a condition satisfied
if the Fermi wavevectors of the two bands have similar magnitude), and becomes long-ranged at a
quantum critical point, there is an extended temperature regime where the
transport time becomes equivalent to the lifetime of the light $c$-electrons.
In the Kondo-Heisenberg
model, as we show below, this gives rise to a quasi-linear $T \log(T)$-dependence.  This is in
contrast with the simple $f$-$c$ scattering model of Fig.~\ref{fig1}(a),
where a $T^{5/3}$-dependence for the transport lifetime
would occur instead~\cite{mathon,maslov}.

\emph{Model.}---
We consider the Kondo-Heisenberg model in three dimensions, whose thermodynamics has been studied
extensively for the physics of Kondo
breakdown at a quantum critical point~\cite{Paul:2007th,Paul:2008bg,Pepin:2007do,Senthil:2003gz}.
It is described by the action
\begin{align}
\label{eq1:action}
S &=
- \frac{1}{\beta} \sum \left[
\bar{c}_{\bk, \mu}( \om_n) G_c^{-1} (\bk, i \om_n) c_{\bk, \mu} ( \om_n)
+ c \rightarrow f \right]
  \nonumber \\
&+
 \frac{1}{\beta} \sum
 \bar{\si}_{\bq}( \Om_n) D^{-1} (\bq, i \Om_n) \si_{\bq}(\Om_n)
 \nonumber \\
&+
\frac{J_K}{\beta^2} \sum
\bar{c}_{\bk, \mu}( \om_n) f_{\bk - \bq, \mu} (\om_n - \Om_m) \si_{\bq} (\Om_m) +{\rm h.c.}
\end{align}
The summations involve all repeated indices. $(c_{\bk, \mu}, f_{\bk, \mu})$ and their conjugates
denote fermion fields
with spin $\mu$ of the two bands.  Their Greens functions can be
written as $G_{a}^{-1} (\bk, i \om_n) = i \om_n - \ep_{\bk}^a + i/(2 \tau_{a}) {\rm sgn}(\om_n)$,
with the index $a=(c,f)$.
The band dispersions are given by $\ep_{\bk}^a \equiv (k^2 - k_{Fa}^2)/(2m_a)$, and the
mass ratio $\al \equiv m_c/m_f \ll 1$ is an important small parameter of the system. The $T$-independent
lifetimes $\tau_c$ and $\tau_f$ are due to impurity scattering, and since they depend inversely on their respective
masses, we have $\tau_f = \al \tau_c$. The Fermi wavevectors are $k_{Fc}$ and $k_{Ff}$ respectively, with
$q^{\ast} \equiv k_{Ff} - k_{Fc}$ denoting the mismatch of the Fermi surfaces.
In the following we take the mismatch to be small, i.e., $(q^{\ast}/k_{Fc}) \ll 1$. This is equivalent to
assuming that the $c$-band is near half-filling, which is not unusual for systems with good metallic
properties such as the heavy fermions. The interaction with hybridization fluctuations
between the bands is mediated by the bosonic fields $(\bar{\si}_{\bq}, \si_{\bq})$. These are the critical modes
of the theory, and close to the quantum critical point their Greens function is
given by~\cite{Paul:2007th,Paul:2008bg}
$D^{-1}(\bq, i\Om_n) = \nu_0 J_K^2 \left[ q^2/(4 k_{Fc}^2) + \pi k_{Fc} \left| \Om_n \right|/(2 \al q \Lambda)
\right]$,
where $\nu_0 \equiv m_c k_{Fc}/(2 \pi^2)$
is the $c$-density of states per spin at the Fermi energy, $J_K$ is the Kondo coupling, and
$\Lambda \equiv k_{Fc}^2/m_c$ is the $c$-bandwidth.
This form of the Greens function is valid above small momentum and
energy cutoffs $q^{\ast}$ and $E^{\ast} \equiv \al \Lambda (q^{\ast}/k_{Fc})^3$ respectively. The
ultraviolet cutoff of the theory is the $f$-bandwidth $\al \Lambda$.

In the following we study the charge transport properties of the model using first a semiclassical
Boltzmann method wherein the issue of the $(1- \cos \theta)$ factor is most transparent. Next we study the
system using a Kubo formalism and establish the equivalence of the particle and transport times. We also
show that the correction to the current vertex as well as the contribution of the collective mode
$\si_{\bq}$, which are not included in the Boltzmann
calculation, are subleading.

\emph{Boltzmann treatment.}---
In the presence of an electric field $\bE$ the Boltzmann equations for the non-equilibrium occupations $f_{\bk}^a$
of the two bands are (setting $\hbar=1$)
\beq
- e \bE \cdot \nabla_{\bk} f_{\bk}^a = - I_{ei}^a \left[ f_{\bk} \right] - I_{ee}^a \left[ f_{\bk} \right],
\nonumber
\eeq
with $a= (c,f)$ for the two bands.
The electron-impurity collision integrals are given by $I_{ei}^a \left[ f_{\bk} \right]
= (f_{\bk}^a - n_{F \bk}^a)/\tau_a$, where $n_{F \bk}^a \equiv 1/[\exp(\ep_{\bk}^a/T) +1]$
are the equilibrium Fermi distributions. The electron-electron collision integral for the
$c$-band can be written as
\begin{align}
I_{ee}^c \left[ f_{\bk} \right] &=
\! 2 J_K^2 \! \sum_{\bq} \! \int_{-\infty}^{\infty} \! d \Om \ {\rm Im}
D (\bq, \Om + i \eta)
\left[ f_{\bk}^c (1- f_{\bk - \bq}^f) \right.
\nonumber \\
& \times \left.
(1 + n_B(\Om)) - (1- f_{\bk}^c) f_{\bk - \bq}^f n_B(\Om) \right]
\nonumber \\
& \times
\delta (\ep_{\bk}^c - \ep_{\bk - \bq}^f - \Om),
\nonumber
\end{align}
with a similar expression for $I_{ee}^f \left[ f_{\bk} \right]$.
$n_{B} (\Om)$ is the equilibrium Bose function.

We solve the above equations in the linear approximation where
$f_{\bk}^a = n_{F \bk}^a - n_{F \bk}^a (1 - n_{F \bk}^a) g_{\bk}^a$
with the ansatz
$
g_{\bk}^a = \left( e \bE \cdot \bv_{\bk}^a \right) t_{a}/T.
$
Here $\bv_{\bk}^a \equiv \nabla_{\bk} \ep_{\bk}^a$ are the band velocities,
and $t_{a}$ are variational parameters to be determined. We get the
simplified equations
\beq
- e\bE \cdot \bv_{\bk}^a \, \bar{n}_{F \bk}^a = T g_{\bk}^a \, \bar{n}_{F \bk}^a / \tau_a
+ \tilde{I}_{ee}^a \left[ g_{\bk} \right],
\nonumber
\eeq
where $\bar{n}_{F \bk}^a \equiv \partial n_{F \bk}^a/\partial \ep_{\bk}^a$, and
\begin{align}
\tilde{I}_{ee}^c \left[ g_{\bk} \right] &=
2 J_K^2 \sum_{\bq} \int_{-\infty}^{\infty} \! d \Om \ {\rm Im}
D (\bq, \Om + i \eta) n_B(\Om) n_{F \bk - \bq}^f
\nonumber \\
& \times
(1 - n_{F \bk}^c) ( g_{\bk}^c  - g_{\bk - \bq}^f )
\delta (\ep_{\bk}^c - \ep_{\bk - \bq}^f - \Om),
\nonumber
\end{align}
with an analogous expression for $\tilde{I}_{ee}^f \left[ g_{\bk} \right]$.
It is important to note that
\beq
\label{eq:g-nonzero}
g_{\bk}^c - g_{\bk - \bq}^f = (e/T)(t_c/m_c - t_f/m_f) (\bE \cdot \bk) + \mathcal{O} (q).
\eeq
In other words, during scattering involving release/absorption of the $\si_{\bq}$-boson,
the velocity imbalance between the outgoing and the incoming fermions
is non-zero even in the limit $q \rightarrow 0$. This ensures that the $(1- \cos \theta)$ factor
does not give rise to any additional $T$-dependence. The crucial ingredient here is the nature of
the interaction, namely the fluctuation of hybridization between a light and a heavy electron,
as opposed to simple $f$-$c$ scattering.

The solution of the Boltzmann equation is standard, and it gives the resistivity
\beq
\label{eq:resistivity1}
\frac{\rho(T)}{\rho_c}
= \frac{1/\tau_c + (1 + \al^2)/\tau_{ee}(T)}{(1+\al^2)/\tau_c + 4\al/\tau_{ee}(T)},
\eeq
where $\rho_c = 3/(2 e^2 \nu_0 v_{Fc}^2 \tau_c)$ is the resistivity of the non-interacting
$c$-subsystem. The interband scattering rate is defined by
\begin{align}
\frac{1}{\tau_{ee}(T)} & \equiv
\frac{6 J_K^2}{\nu_0 k_{Fc}^2 T} \sum_{\bk \bq}
\int_{-\infty}^{\infty} d \Om \ {\rm Im} D (\bq, \Om + i \eta) n_B(\Om)
\nonumber \\
& \times
(\bk \cdot \hat{e})^2
 n_{F \bk - \bq}^f (1 - n_{F \bk}^c)
\delta (\ep_{\bk}^c - \ep_{\bk - \bq}^f - \Om),
\nonumber
\end{align}
where $\hat{e}$ defines the direction of the electric field.
For $T > E^{\ast}$ we find
\beq
\label{eq:resistivity-ee}
\tau_{ee}(T)^{-1} = 2 k_{Fc}^3/(3 \pi \nu_0) (T/\al \Lambda) \ln (T/E^{\ast}),
\eeq
while for $T< E^{\ast}$ the singularity of the interaction is cut off and we get the
regular Fermi liquid $T^2$-dependence. Furthermore, since the electron masses are
very different, we are guaranteed a rather extended temperature regime, defined by
$1/\tau_c \ll 1/\tau_{ee} \ll 1/(\al \tau_c)$, where the resistivity has a quasi-linear
temperature dependence with
\beq
\label{eq:resistivity-linear}
 \rho(T) = \rho_c \tau_c/\tau_{ee}(T).
\eeq
This equation, with quasi-linear temperature dependence of the resistivity,
is interesting from the point of view of the phenomenology of the
heavy fermions near quantum criticality.

\begin{figure}[!!t]
\begin{center}
\includegraphics[width=5cm]{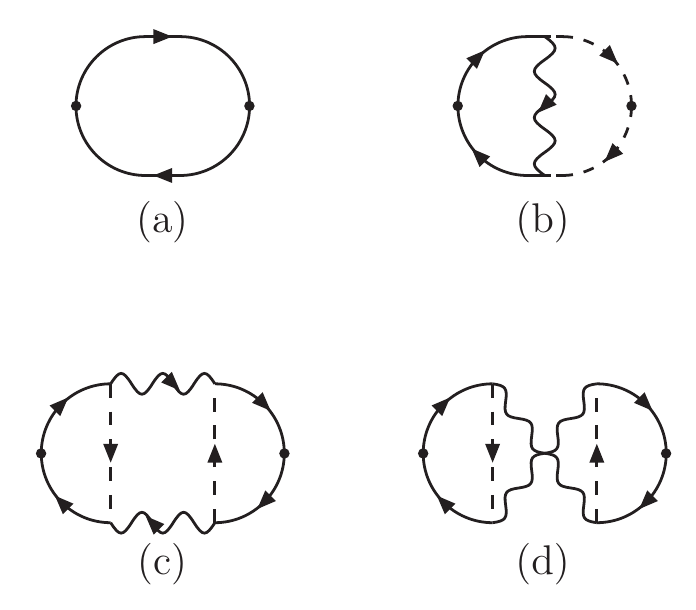}
\caption{
(a)-(c). Graphs for the current-current correlation function used for the Kubo
calculation of the conductivity. (d) is not a valid graph in the current theory.
}
\label{fig2}
\end{center}
\end{figure}

\emph{Kubo treatment.}--- We show that the above result is the same as
the resistivity of the $c$-band with its particle lifetime taken as the transport time.
The self energy of the $c$-electrons, defined by
\beq
\Sigma_c (\bp, i \om_n) =
(J_K^2/\beta) \sum_{\nu_n, \bq} D(\bq, i \nu_n)
G_f (\bp + \bq, i \om_n - i \nu_n),
\nonumber
\eeq
has been evaluated earlier~\cite{Paul:2007th,Paul:2008bg}. In particular,
it has been shown that the hybridization fluctuation generates marginal Fermi
liquid properties with
${\rm Im} \Sigma_c^R (k_{Fc}, \om) \propto - |\om|/\alpha$ for $|\om| > E^{\ast}$.
At finite temperature this
translates into a quasiparticle lifetime
$\tau_{qp}(T)^{-1} \equiv -
2 {\rm Im} \Sigma_c^R $,
with
\beq
\label{eq:quasiparticle-time}
\tau_{qp}(T, \om =0)^{-1} = 2 k_{Fc}^3/(3 \pi \nu_0) (T/\al \Lambda)
\ln (T/E^{\ast})
\eeq
for $T > E^{\ast}$, and a regular
$T^2$-dependence below the cutoff. Comparing with
Eq.~(\ref{eq:resistivity-ee}) we find that the $c$-quasiparticle
lifetime $\tau_{qp}$ and the transport lifetime $\tau_{ee}$ are the
same.
Indeed, when we calculate the conductivity using
the simplest Kubo bubble without vertex corrections
(Fig.~\ref{fig2}(a)), which is the same as identifying $\tau_{qp}$
as the transport lifetime, we get
\beq
\label{eq:Kubo-resistivity}
\rho(T)_{\rm Kubo} = \rho_c \tau_c/ \tau_{qp}(T),
\eeq for
$1/\tau_{qp} \gg 1/\tau_c$. This
result is the same as the one obtained in Eq.~(\ref{eq:resistivity-linear})
from the Boltzmann treatment
where the issue of the $(1- \cos \theta)$ factor has been dealt with
explicitly.
This completes the demonstration that in a two-band
model of light and heavy electrons, the quasiparticle lifetime
obtained from critical hybridization fluctuations is quasi-linear in
temperature and is equal to the transport lifetime.

Next, in order to verify that indeed the leading temperature dependence of the resistivity
is obtained from the simplest Kubo bubble, we examine two higher order graphs and show that
they give subleading contributions.
(a) We calculate the contribution of the graph with a single vertex
correction (Fig~\ref{fig2}(b))
which can be written as
\begin{align}
\sigma_V &= -{\rm Im} \sum_{\bp} \bv_p^c \int_{- \infty}^{\infty} \frac{d \om}{ 4 \pi i}
\frac{\ptl}{\ptl \om} \tanh [\om/(2T)] G_c^R (\bp, \om)
\nonumber \\
& \times
 G_c^A (\bp, \om) \bL_c (\om + i \eta, \om - i \eta, \bp),
\nonumber
\end{align}
in the approximation where the dispersions can be linearized.
Here the vertex is defined as
\begin{align}
\bL_c (i \om_{n1}, i \om_{n2}, \bp)
&\equiv \frac{J_K^2}{\beta} \sum_{\nu_n, \bq} \bv_{\bp + \bq}^f
G_f (\bp + \bq, i \om_{n1} - i \nu_n)
\nonumber \\
& \times
G_f (\bp + \bq, i \om_{n2} - i \nu_n)
D(\bq, i \nu_n).
\nonumber
\end{align}
At $T=0$ we get $\bL_c \propto (k_{Fc}^2 \tau_f \om/\nu_0) \hat{p}$, where $\hat{p}$ is the
unit vector along $\bp$. We find that in the relevant temperature regime $v_{Fc} \gg | \bL_c |$,
which guarantees that $\sigma_V$ is subleading. This conclusion is also consistent with the general
picture that the current is effectively relaxed by the same processes that give rise to the quasiparticle
lifetime, and therefore one does not expect the vertex corrections to play a crucial role.
(b) We next calculate the Azlamasov-Larkin (AL) graph which can also be interpreted as the contribution
of the boson mode to the conductivity (Fig.~\ref{fig2}(c)).
This contribution was not taken into account in the
semiclassical Boltzmann treatment.
Note that, unlike in single band models where there are two
non-equivalent AL graphs, in the current model there is only a single
one (compare Fig.~\ref{fig2}(c) \& (d)). We find that the
leading contribution can be written as
\begin{align}
\sigma_{AL} &= -  \sum_{\bq} \int_{- \infty}^{\infty} \frac{d \om}{ 4 \pi }
\frac{\ptl}{\ptl \om} \coth [\om/(2T)] \left| D^R(\bq, \om) \right|^2
\nonumber \\
& \times
\bL_{\si} (\om + i \eta, \om - i \eta, \bq)^2,
\nonumber
\end{align}
where the boson vertex at $T=0$ is
\beq
\bL_{\si} \approx J_K^2 \ {\rm Im} \sum_{\bp} \bv_{\bp}^c \int_0^{\om} \frac{d \nu}{\pi}
\left| G_c^R(\bp, \nu) \right|^2  G_f^R (\bp - \bq, \nu - \om).
\nonumber
\eeq
All other contributions to $\si_{AL}$, including those from the static vertex
$\bL_{\si}(\om =0)$, are subleading in the relevant temperature regime. We find,
$\bL_{\si} \propto - (\nu_0 \tau_c \om^2)/(v_{Fc} \al^2 q^2)\hat{q}$, from which
we estimate $\si_{AL} \propto v_{Fc}^2 \tau_c^2 k_{Fc}^3 (T/\alpha\Lambda)^{7/3}$.
This implies that at low enough temperature
$T < T_{AL} \equiv \al \Lambda^{2/5}/\tau_c^{3/5}$ the AL contribution is subleading.

\emph{Discussion.}---
Thus, from the Boltzmann and the Kubo calculations,
we can conclude that in the current model the resistivity has a leading
$T \log T$ dependence over a temperature regime defined by
${\rm max}(E^{\ast}, \al/\tau_c) < T < {\rm min}( 1/\tau_c, T_{AL})$.
In this regime the conductivity is essentially due to the $c$-subsystem
with the $f$-subsystem providing a bath for the current to relax.
Furthermore, from the solution of the Boltzmann equation we find that
in Eq.~(\ref{eq:g-nonzero})
$g_{\bk - \bq}^f \approx 0$ to leading order in $\alpha$. This implies that,
to leading order, the current theory maps to a model of ``impurity" scattering with a
potential that is temperature dependent, but is independent of the scattering angle
of the light fermions.
This completes the physical picture why the transport and the quasiparticle lifetimes are
equivalent in our model.

An important aspect of the Kondo-Heisenberg model is that the half-filled
correlated $f$-subsystem should have one fermion per site in its physical
Hilbert space. From the point of view of its dynamics, this translates into the
constraint $\bJ_f - \bJ_{\si} =0$ for the currents of the $f$- and the $\si$-
subsystems. This implies that even in the gauge where the $f$-electrons are
charged (which is what we have implicitly assumed), there is a backflow contribution
from the $\si$-bosons. This contribution is explicitly ignored in our Boltzmann
treatment, where the boson distribution is taken to be at equilibrium. However,
since in the relevant $T$-regime the conductivity is entirely due to the
$c$-subsystem, and the role of the $f$-fermions and its intrinsic scattering rate is merely
to provide a high temperature cutoff to this regime, we expect that taking the
constraint into account will only modify this high-$T$ cutoff, and not change the main
content of our result. This expectation is
also consistent with the finding in the Kubo formalism that the AL graph, which is the
boson contribution to the conductivity,
only provides a high temperature cutoff to the $T \log T$ behaviour.

In summary, we demonstrated that the Kondo-Heisenberg model gives rises to a novel $T \log T$ behaviour
for both the single-particle and transport lifetimes, which appears to be consistent with data in
several heavy fermion systems.  The equivalence of the two lifetimes is a consequence of
the transmutational nature of the $f$-$c$ hybridization, and may be relevant to other
mutli-band theories for correlated electrons.

We are very thankful to Dmitrii Maslov and Andrey Chubukov for
several illuminating
discussions that were crucial for this work.
Work at Argonne was supported by the U.~S.~Dept.~of Energy, Office of Science, Basic Energy
Sciences, under contract DE-AC02-06CH11357.
I.~P.~ and C.~P.~ would like to thank Argonne staff for their hospitality during their respective visits.

\end{document}